\documentclass[aps,eqsecnum,preprint,floats,epsf,epsfig,nofootinbib,showpacs]{revtex4}
\usepackage{graphicx}
\def\be{\begin{eqnarray}}
\def\en{\end{eqnarray}}
\def\non{\nonumber}
\def\la{\langle}
\def\ra{\rangle}

\def\prl{{ Phys. Rev. Lett.}~}

\def\bi{\bibitem}

\begin{document}

\title{\Large \bf Leading-twist light cone distribution amplitudes
for $p$-wave heavy quarkonium states
 }

\author{ \bf  Chien-Wen Hwang\footnote{
t2732@nknucc.nknu.edu.tw}}

\affiliation{\centerline{Department of Physics, National Kaohsiung Normal University,} \\
\centerline{Kaohsiung, Taiwan 824, Republic of China}
 }


\begin{abstract}
In this paper, a study of light-cone distribution amplitudes for
$p$-wave heavy quarkonium states are presented. Within the
light-front framework, the leading twist light-cone distribution
amplitudes, and their relevant decay constants, have some simple
relations. These relations can be further simplified when the
non-relativistic limit and the wave function as a function of
relative momentum $|\vec\kappa|$ are taken into consideration. In
addition, the $\kappa_\perp$ integrations in the equations of LCDAs
and $\xi$-moments can be completed analytically when the
Gaussian-type wave function is considered. After fixing the
parameters that appear in the wave function, the curves and the
corresponding decay constants of the LCDAs are plotted and
calculated for the charmonium and bottomonium states. The first
three non-vanishing $\xi$-moments of the LCDAs are estimated and are
consistent with those of other theoretical approaches.
\end{abstract}
\maketitle %

\section{Introduction}
Light-cone distribution amplitudes (LCDAs) of hadrons are key
ingredients in the description of various exclusive processes of
quantum Chromodynamics (QCD), and their role can be analogous to
those of parton distributions in inclusive processes. In terms of
the Bethe-Salpeter wave functions $\Phi(u_i,k_{i\perp})$, LCDAs
$\phi(u_i)$ are defined by retaining the momentum fractions $u_i$
and integrating out the transverse momenta $k_{i\perp}$ \cite{LB}.
They provide essential information on the non-perturbative structure
of the hadron for QCD treatment of exclusive reactions.
Specifically, the leading twist LCDAs describe the probability
amplitudes to find the hadron in a Fock state with the minimum
number of constituents. In addition, the fact that $B$-physics
exclusive processes are under investigation in BaBar and Belle
experiments, also urges the detailed study of hadronic LCDAs. In
literature, there have been many non-perturbative approaches to
estimate LCDAs, such as the QCD sum rules
\cite{CZ,Bakulev,Ball1,Yang1,Bragutanew}, lattice calculation
\cite{Ali,Braun}, chiral quark model from the instanton vacuum
\cite{Petrov,Nam}, Nabmbu-Jona-Lasinio model
\cite{Arriola,Praszalowicz}, and the light-front quark model
\cite{Ji1,me,Ji2}. These studies have dealt with LCDAs of
pseudoscalar
\cite{Bakulev,Braun,Petrov,Nam,Arriola,Praszalowicz,Ji1,me}, vector
\cite{Ball1,Ali,Ji1,me}, axial vector \cite{Yang1,Bragutanew,Ji2},
and tensor \cite{Bragutanew} mesons.

The present paper is devoted to the study of leading twist LCDAs of
$p$-wave heavy quarkonium states which include the scalar
($\chi_{c0}, \chi_{b0}$), axial vector ($\chi_{c1}, \chi_{b1}, h_c,
h_b$), and tensor ($\chi_{c2}, \chi_{b2}$) mesons. The motivation of
this study is as follows. Since the discoveries of $J/\psi$ and
$\Upsilon$, occurring more than thirty years ago, a great deal of
information on heavy quarkonium levels and their transitions has
been accumulated \cite{PDG08}. The numerous transitions between
heavy quarkonium states are classified as strong and radiative
decays, which shed light on aspects of QCD in both the perturbative
and the non-perturbative regimes (for a recent review see
\cite{EGMR}). In particular, some experimental results regarding
$\chi_{cJ}$ mesons have recently been reported
\cite{CLEO1,CLEO2,CLEO3,CLEO4}. Therefore, a thorough understanding
of their properties, such as LCDAs which are the universal
non-perturbative objects, will be of great benefit when analyzing
the hard exclusive processes with heavy quarkonium production.

It is known that heavy quarkonium is relevant to non-relativistic
treatments \cite{QR}. Although non-relativistic QCD (NRQCD) is a
powerful theoretical tool for separating high-energy modes from
low-energy contributions, in most cases the calculation of
low-energy hadronic matrix elements has relied on model-dependent
non-perturbative methods. In this study, heavy quarkonium is
explored within the light-front quark model (LFQM) which is a
promising analytic method for solving the non-perturbative problems
of hadron physics \cite{BPP} as well as offers many insights into
the internal structures of bound states. The basic ingredient in
LFQM is the relativistic hadron wave function which generalizes
distribution amplitudes by including transverse momentum
distributions, and contains all the information of a hadron from its
constituents. The hadronic quantities are represented by the overlap
of wave functions and can be derived in principle. The light-front
wave function is manifestly a Lorentz invariant, expressed in terms
of internal momentum fraction variables which are independent of the
total hadron momentum. Moreover, the fully relativistic treatment of
quark spins and center-of-mass motion can be carried out using the
so-called Melosh rotation \cite{LFQM}. This treatment has been
successfully applied to calculate phenomenologically many important
meson decay constants and hadronic form factors \cite{Jaus1, CCH1,
Jaus2, CCH2, Hwang, Wei}. Therefore, the main purpose of this study
is the calculation of the leading twist LCDAs of $p$-wave heavy
quarkonium states within LFQM.

The remainder of this paper is organized as follows. In Section II,
the leading twist LCDAs of $p$-wave meson states are shown in cases
of the vector and tensor currents. In Section III, the formulism of
LFQM is reviewed briefly, then the leading twist LCDAs are extracted
within LFQM. The $\xi$-moments of these LCDAs are also calculated.
In Section IV, numerical results and discussions are presented.
Finally, the conclusions are given in Section V.

\section{Leading twist LCDAs of $p$-wave mesons}
Amplitudes of hard processes involving $p$-wave mesons can be
described by the matrix elements of gauge-invariant nonlocal
operators, which are sandwiched between the vacuum and the meson
states,
 \be
 \langle 0 |\bar q (x) \Gamma [x,-x] q(-x) | H(P,\epsilon)\rangle,
 \label{nonlocal}
 \en
where $P$ is the meson momentum, $\epsilon$ is the polarization
vector or tensor (of course, $\epsilon$ does not exist in the case
of scalar meson), $\Gamma$ is a generic notation for the Dirac
matrix structure and the path-ordered gauge factor is:
 \be
 [x,y]={\textrm{P exp}}\left[ig_s\int^1_0 dt(x-y)_\mu
 A^\mu(tx+(1-t)y)\right].
 \en
This factor is equal to unity in the light-cone gauge which is
equivalent to the fixed-point gauge, $(x-y)_\mu A^\mu (x-y)=0$, as
the quark-antiquark pair is at the light-like separation
\cite{Yang2}. For simplicity, the gauge factor will not be shown
below.

The asymptotic expansion of exclusive amplitudes, in powers of large
momentum transfer, is governed by the expanding amplitude Eq.
(\ref{nonlocal}), shown in powers of deviation from the light-cone
$x^2=0$. There are two light-like vectors, $p$ and $z$, which can be
introduced by:
 \be
 p^2=0,~~~~~~z^2=0,
 \en
such that $p \to P$ in the limit $M_H^2 \to 0$ and $z \to x$ for
$x^2 = 0$. From this it follows that \cite{Ball1}
 \be
 z^\mu &=& x^\mu - P^\mu \frac{1}{M_H^2} \left[Px-\sqrt{(Px)^2-x^2
 M_H^2}\right] \non \\
 &=& x^\mu - P^\mu \frac{x^2}{2 P z}+ O(x^4), \non \\
  p^\mu &=& P^\mu -z^\mu \frac{M^2_H}{2 P z},\label{Pp}
 \en
where $P x \equiv P \cdot x$ and $P z=p z=\sqrt{(P x)^2-x^2 M_H^2}$.
In addition, if the meson is assumed that it moves in the positive
$\hat{e}_3$ direction, then $p^+$ and $z^-$ are the only nonzero
component of $p$ and $z$, respectively, in an infinite momentum
frame. For the axial vector meson, the polarization vector
$\epsilon^\mu$ is decomposed into longitudinal and transverse
projections as:
 \be
 \epsilon^\mu_{\|}=\frac{\epsilon z}{p z}\left(p^\mu-z^\mu\frac{M^2_H}{2p z}\right),
 ~~~\epsilon^\mu_{\perp}=\epsilon^\mu-\epsilon^\mu_{\|}, \label{epsilon}
 \en
respectively. For the tensor meson, the polarization tensor is:
 \be
 \epsilon^{\mu\nu}(m)=\langle 11;m'm''|11;2m \rangle \epsilon^\mu (m')
 \epsilon^\nu (m''),
 \en
($m$ is the magnetic quantum number) and $\epsilon_{\mu\bullet}
(\equiv \epsilon^{\mu\nu} z_\nu)$ can also be decomposed into
longitudinal and transverse projections as:
 \be
 \epsilon^{\mu\bullet}_\|=\frac{\epsilon^{\bullet\bullet}}{p z}\left(p^\mu-z^\mu\frac{M^2_H}{2p z}\right),
 ~~~\epsilon^{\mu\bullet}_\perp=\epsilon^{\mu\bullet}-\epsilon^{\mu\bullet}_\|.
 \en

LCDAs are defined in terms of matrix element of nonlocal operator in
Eq. (\ref{nonlocal}). For the scalar $(S)$, axial vector $(A)$, and
tensor $(T)$ mesons, the leading twist LCDAs can be defined as:
 \be
 \langle 0|\bar q (z) \gamma^\mu q (-z)|S(P)\rangle &=& f_S \int^1_0 du~e^{i\xi p z}\left[p^\mu
  \phi_S(u)+z^\mu \frac{M_S^2}{2p z}g_S(u)\right], \label{S}\\
 \langle 0|\bar q (z) \gamma^\mu \gamma_5 q (-z)|A(P,\epsilon_{\lambda=0})\rangle
 &=&if_A M_A \int^1_0 du~e^{i\xi p z}\Big\{p^\mu \frac{\epsilon z}{p z}
 \phi_{A\|}
 (u) +\epsilon^\mu_{\perp}g_{A\perp}(u)\non \\
 &&~~~~~~~~~~~~~~~~~~~~~~- z^\mu \frac{\epsilon z}{2 (p z)^2}M^2_A
 g_{A3}(u)\Big\}, \label{AL}
 \en
 \be
 &&\langle 0|\bar q (z) \sigma^{\mu\nu} \gamma_5 q (-z)|A(P,\epsilon_{\lambda=\pm 1})\rangle
 =f^\perp_A \int^1_0 du~e^{i\xi p z}\Big\{(\epsilon^\mu_{\perp} p^\nu-\epsilon^\nu_{\perp} p^\mu)
 \phi_{A\perp} (u)\non \\
 &&\qquad\qquad\qquad\qquad\qquad\qquad+(p^\mu z^\nu-p^\nu z^\mu)\frac{M^2_A \epsilon z}{(p z)^2}h_{A\|}(u)
 +(\epsilon^\mu_{\perp} z^\nu-\epsilon^\nu_{\perp} z^\mu) \frac{M^2_A}{2 p z}
 h_{A3}(u)\Big\},\label{AT}\\
 &&\langle 0|\bar q (z) \gamma^\mu q (-z)|T(P,\epsilon_{\lambda=0})\rangle
 =f_T M_T^2 \int^1_0 du~e^{i\xi p z}\Big\{p^\mu \frac{\epsilon^{\bullet\bullet}}{(p z)^2}
 \phi_{T\|}(u) +\frac{\epsilon^{\mu\bullet}_\perp}{p z}g_{T\perp}(u)\non \\
 &&\qquad\qquad\qquad\qquad\qquad\qquad- z^\mu \frac{\epsilon^{\bullet\bullet}}
 {2 (p z)^3}M^2_T g_{T3}(u)\Big\},\label{TL} \\
 &&\langle 0|\bar q (z) \sigma^{\mu\nu} q (-z)|T(P,\epsilon_{\lambda=\pm 1})\rangle
 =if^\perp_T M_T \int^1_0 du~e^{i\xi p
 z}\Big\{\frac{(\epsilon^{\mu\bullet}_\perp
 p^\nu-\epsilon^{\bullet\nu}_\perp
 p^\mu)}{p z}
 \phi_{T\perp} (u)\non \\
 &&\qquad\qquad\qquad\qquad\qquad+(p^\mu z^\nu-p^\nu z^\mu)\frac{M^2_T
 \epsilon^{\bullet\bullet}}{(p z)^3}h_{T\|}(u)
 +(\epsilon^{\mu\bullet}_\perp z^\nu-\epsilon^{\bullet\nu}_\perp z^\mu) \frac{M^2_T}{2 (p z)^2}
 h_{T3}(u)\Big\},\label{TT}
 \en
where $u$ is the momentum fraction and $\xi \equiv (1-u)-u =1- 2 u$.
Here $\phi_S$, $\phi_{A,T\|}$, and $\phi_{A,T\perp}$ are the leading
twist-$2$ LCDAs, and the others contain contributions from
higher-twist operators. Due to G-parity, $\phi_S$, $g_S$,
$\phi_{^3A_1 \perp}$, $h_{^3A_1\|}$, $h_{^3A_1 3}$,
$\phi_{^1A_1\|}$, $g_{^1A_1\perp}$, $g_{^1A_1 3}$, $\phi_{T\|}$,
$g_{T\perp}$, $g_{T3}$, $\phi_{T\perp}$, $h_{T\|}$, and $h_{T3}$ are
antisymmetric (odd) under replacement $u \to 1-u$, whereas,
$\phi_{^1A_1 \perp}$, $h_{^1A_1\|}$, $h_{^1A_1 3}$,
$\phi_{^3A_1\|}$, $g_{^3A_1\perp}$, and $g_{^3A_1 3}$ are symmetric
(even) in the quarkonium states. Therefore, the leading twist LCDAs
are normalized as:
 \be
 \int^1_0 du \xi \phi^{({\rm odd})} (u) = 1, ~~~~~\int^1_0 du  \phi^{(\rm even)} (u) =
 1. \label{normal}
 \en
and can be expanded \cite{CZ} in Gegenbauer polynomials
$C^{3/2}_n(\xi)$ as
 \be
 \phi(\xi,\mu)= \phi_{\rm as}(\xi)\left[\sum_{l=0}^\infty
 a_l(\mu)C_l^{3/2}(\xi)\right].
 \en
where $\phi_{\rm as}(\xi)=3(1-\xi^2)/4$ is the asymptotic quark
distribution amplitude and $a_l(\mu)$ are the Gegenbauer moments
which describe to what degree the quark distribution amplitude
deviates from the asymptotic one. $C^{3/2}_l(\xi)$s have the
orthogonality integrals
 \be
 \int^1_{-1} (1-\xi^2)C^{3/2}_l(\xi) C^{3/2}_m (\xi) d\xi =
 \frac{2(l+1)(l+2)}{2l+3}~\delta_{lm}. \label{orthogonal}
 \en
Then $a_l$ can be obtained by using the above orthogonality
integrals as
 \be
 a_l(\mu) &=&\frac{2(2
 l+3)}{3(l+1)(l+2)}\int^1_{-1}C^{3/2}_l(\xi)\phi(\xi,\mu)d\xi.
 \en
An alternative approach to parameterize quark distribution amplitude
is to calculate the so-called $\xi$-moments
 \be
 \langle \xi^n\rangle_\mu=\int^1_{-1} d\xi~\xi^n \phi(\xi,\mu).
 \en

To disentangle the twist-$2$ LCDAs from higher twist in Eqs.
(\ref{S}) $\sim$ (\ref{TT}), the twist-$2$ contribution of the
relevant nonlocal operator $\bar q (z) \Gamma  q(-z)$ must be
derived. For the $\Gamma = \gamma^\mu (\gamma_5)$ case, the leading
twist-$2$ contribution contains contributions of the operators which
are fully symmetric in Lorentz indices \cite{Ball2,BB}:
 \be
 [\bar q (-z)\gamma^\mu (\gamma_5)
 q(z)]_2=\sum^\infty_{n=0} \frac{1}{n!}\bar q
 (0)\bigg\{\frac{(z\cdot \widehat{D})^n}{n+1} \gamma^\mu +
 \frac{n (z\cdot \widehat{D})^{n-1}}{n+1}\widehat{D}^\mu {\not
 \!z}\bigg\}(\gamma_5)q(0), \label{twist2expand}
 \en
where $\widehat{D}=\overrightarrow{D}-\overleftarrow{D}$ and
$\overrightarrow{D}=\overrightarrow{\partial}-ig B^a (\lambda^a/2)$.
The sum can be repressed in terms of a nonlocal operator,
 \be
 [\bar q (-z)\gamma^\mu (\gamma_5)
 q(z)]_2= \int^1_0 dt \frac{\partial}{\partial z_\mu} \bar q (-t z) \not
 \!z (\gamma_5) q(t z). \label{t1}
 \en
Taking the matrix element between the vacuum and the $p$-wave meson
state, we obtain:
 \be
 \langle 0 |[\bar q (-z)\gamma^\mu
 q(z)]_2|S(P)\rangle &=& f_S \int^1_0 du
 \phi_S(u) \Bigg\{p^\mu e^{i\xi p z} +(P^\mu -p^\mu)\int^1_0 dte^{i\xi
 t p z}\Bigg\}, \label{Sphi}\\
 \langle 0 |[\bar q (-z)\gamma^\mu \gamma_5
 q(z)]_2|A(P,\epsilon_{\lambda=0})\rangle &=& if_A M_A \int^1_0 du
 \phi_{A\|}(u) \Bigg\{p^\mu \frac{\epsilon z}{p z} e^{i\xi p z}\non
 \\
 &&~~~~~~~~~~~~~~~~~~+\left(\epsilon^\mu-p^\mu \frac{\epsilon z}{p z}\right)\int^1_0 dte^{i\xi
 t p z}\Bigg\},\label{Aphi}\\
 \langle 0 |[\bar q (-z)\gamma^\mu
 q(z)]_2|T(P,\epsilon_{\lambda=0})\rangle &=& f_T M^2_T \int^1_0 du
 \phi_{T\|}(u) \Bigg\{p^\mu \frac{\epsilon^{\bullet\bullet}}{(p z)^2} e^{i\xi p
 z}\non \\
 &&~~~~~~~~~~~~~~~~~~+2\left(\frac{\epsilon^{\mu\bullet}}{p z}-p^\mu
 \frac{\epsilon^{\bullet\bullet}}{(p z)^2}\right)\int^1_0 dte^{i\xi
 t p z}\Bigg\}. \label{Tphi}
 \en
The derivations of Eqs. (\ref{Sphi}) $\sim$ (\ref{Tphi}) are shown
in Appendix A. We can use Eq. (\ref{twist2expand}), and then expand
the right-hand sides of Eqs. (\ref{Sphi}) $\sim$ (\ref{Tphi}), as
 \be
 &&~~~\sum^\infty_{n=0} \frac{1}{n!}\langle 0|\bar q
 (0)\bigg\{\frac{(z\cdot \widehat{D})^n}{n+1} \gamma^\mu +
 \frac{n (z\cdot \widehat{D})^{n-1}}{n+1}\widehat{D}^\mu {\not
 \!z}\bigg\}q(0)|S(P)\rangle\non \\
 &=& f_S \sum^\infty_{n=0}
 \frac{i^n}{n!}\int^1_0 du \phi_S(u) (\xi p z)^n \Bigg\{p^\mu +(P^\mu -p^\mu)\int^1_0
 dt t^n\Bigg\},\label{Sn}\\
 &&~~~\sum^\infty_{n=0} \frac{1}{n!}\langle 0|\bar q
 (0)\bigg\{\frac{(z\cdot \widehat{D})^n}{n+1} \gamma^\mu +
 \frac{n (z\cdot \widehat{D})^{n-1}}{n+1}\widehat{D}^\mu {\not
 \!z}\bigg\}q(0)|A(P,\epsilon)\rangle\non \\
 &=& if_A M_A \sum^\infty_{n=0}
 \frac{i^n}{n!}\int^1_0 du \phi_{A\|}(u) (\xi p z)^n \Bigg\{p^\mu \frac{\epsilon z}{p z}
  +\left(\epsilon^\mu -p^\mu \frac{\epsilon z}{p z}\right)\int^1_0
 dt t^n\Bigg\},\label{An}\\
 &&~~~\sum^\infty_{n=0} \frac{1}{n!}\langle 0|\bar q
 (0)\bigg\{\frac{(z\cdot \widehat{D})^n}{n+1} \gamma^\mu +
 \frac{n (z\cdot \widehat{D})^{n-1}}{n+1}\widehat{D}^\mu {\not
 \!z}\bigg\}q(0)|T(P,\epsilon)\rangle\non \\
 &=& f_T M^2_T \sum^\infty_{n=0}
 \frac{i^n}{n!}\int^1_0 du \phi_{T\|}(u) (\xi p z)^n \Bigg\{p^\mu \frac{\epsilon^{\bullet\bullet}}{(p z)^2}
  +2\left(\frac{\epsilon^{\mu\bullet}}{p z} -p^\mu \frac{\epsilon^{\bullet\bullet}}{(p z)^2}\right)\int^1_0
 dt t^n\Bigg\}, \label{Tn}
 \en
respectively. Picking $n=0$ in Eqs. (\ref{Sn}) and (\ref{An}), we
obtain
 \be
 \langle 0|\bar q (0) \gamma^\mu q(0)|S(P)\rangle &=& f_S P^\mu
 \int^1_0 du \phi_S (u), \label{S0}\\
 \langle 0|\bar q (0) \gamma^\mu q(0)|A(P,\epsilon_{\lambda=0})\rangle &=& i f_A M_A \epsilon^\mu
 \int^1_0 du \phi_{A\|} (u).\label{A0}
 \en
Note that the tensor meson cannot be produced by the $V-A$ current,
then we pick $n=1$ in Eq. (\ref{Tn}) and obtain
 \be
 \frac{1}{2}\langle 0|\bar q (0) (\gamma^\mu
 z\cdot\widehat{D}
 +\not\!z \widehat{D}^\mu) q(0)|T(P,\epsilon_{\lambda=0})\rangle &=& f_T M_T^2
 \epsilon^{\mu\bullet} \int^1_0 du \xi\phi_{T\|} (u).\label{T1}
 \en
From the normalization Eq. (\ref{normal}), we have $\langle 0|\bar q
\gamma^\mu \gamma_5 q|^3A_1(P,\epsilon)\rangle = i f_{^3A_1}
 M_{^3A_1} \epsilon^\mu$ which is consistent with the results of \cite{Yang3}.

Next, we consider the case of $\Gamma = \sigma_{\mu\nu}(\gamma_5)$,
where the leading twist-$2$ contribution contains contributions of
the operators:
 \be
 [\bar q (-z)\sigma^{\mu\nu} (\gamma_5)
 q(z)]_2&=&\sum^\infty_{n=0} \frac{1}{n!}\bar q
 (0)\bigg\{\frac{(z\cdot \widehat{D})^n}{2 n+1} \sigma^{\mu\nu} +
 \frac{n (z\cdot \widehat{D})^{n-1}}{2 n+1}\widehat{D}^\mu
 \sigma^{\bullet\nu}\non \\
 &&\qquad\qquad\quad+\frac{n (z\cdot \widehat{D})^{n-1}}{2
 n+1}\widehat{D}^\nu
 \sigma^{\mu\bullet}\bigg\}(\gamma_5)q(0). \label{twist2expandsigma}
 \en
The sum can be also represented in terms of nonlocal operators:
 \be
 [\bar q (-z)\sigma^{\mu\nu} (\gamma_5) q(z)]_2= \int^1_0 dt \left[\frac{\partial}
 {\partial z_\mu} \bar q (-t^2 z) \sigma^{\bullet\nu} (\gamma_5) q(t^2 z)+z_\alpha \frac{\partial}
 {\partial z_\nu} \bar q (-t^2 z) \sigma^{\mu\alpha} (\gamma_5) q(t^2 z)\right]. \label{t2}
 \en
Taking the matrix element between the vacuum and the axial-vector
and tensor meson state, we obtain:
 \be
 &&\langle 0 |[\bar q (-z)\sigma^{\mu\nu} \gamma_5
 q(z)]_2|A(P,\epsilon_{\lambda=\pm 1})\rangle \non \\
 &=& f^\perp_A \int^1_0 du
 \Bigg\{\phi_{A\perp}(u) \bigg[{\cal S^{\mu\nu}}  e^{i\xi p z}
 +\bigg((\epsilon^\mu P^\nu-\epsilon^\nu P^\mu)-
 {\cal S^{\mu\nu}}\bigg)\int^1_0 dte^{i\xi t^2 p z}\bigg]\non \\
 &&\quad\quad\qquad+\bigg(h_{A\|}(u)-\phi_{A\perp}(u)\bigg)
 \Bigg[{\cal T^{\mu\nu}} e^{i\xi p z}+\bigg({\cal U^{\mu\nu}}-
 {\cal T^{\mu\nu}}\bigg)\int^1_0 dte^{i\xi t^2 p
 z}\Bigg]\Bigg\},\label{Aphisigma}\\
 &&\langle 0 |[\bar q (-z)\sigma^{\mu\nu}
 q(z)]_2|T(P,\epsilon_{\lambda=\pm 1})\rangle \non \\
 &=& if^\perp_T M_T \int^1_0 du
 \Bigg\{\phi_{T\perp}(u) \bigg[{\cal S'^{\mu\nu}}  e^{i\xi p z}
 +\bigg(\frac{2(\epsilon^{\mu\bullet} P^\nu-\epsilon^{\nu\bullet} P^\mu)}{p z}-
 3{\cal S'^{\mu\nu}}\bigg)\int^1_0 dte^{i\xi t^2 p z}\bigg]\non \\
 &&\quad\quad\qquad+\bigg(h_{T\|}(u)-\phi_{T\perp}(u)\bigg)
 \Bigg[{\cal T'^{\mu\nu}} e^{i\xi p z}+\bigg(\frac{2{\cal U'^{\mu\nu}}}{p z}-3
 {\cal T'^{\mu\nu}}\bigg)\int^1_0 dte^{i\xi t^2 p
 z}\Bigg]\Bigg\},\label{Tphisigma}
 \en
where
 \be
 {\cal S^{\mu\nu}}&=&\frac{1}{2}\Bigg[(\epsilon^\mu P^\nu-\epsilon^\nu
 P^\mu)
 -(\epsilon^{\mu}_\perp z^\nu -\epsilon^{\nu}_\perp z^\mu)\frac{M_A^2}{2 p z}\Bigg],\non \\
 {\cal T^{\mu\nu}}&=&\frac{\epsilon z M_A^2}{2 (p z)^2} (p^\mu z^\nu-p^\nu z^\mu)
 ,\qquad\qquad
 {\cal U^{\mu\nu}}=\frac{M_A^2}{p z} (\epsilon^\mu z^\nu-\epsilon^\nu z^\mu)
 ,\non \\
 {\cal S'^{\mu\nu}}&=&\frac{1}{2 p z}\Bigg[(\epsilon^{\mu\bullet} P^\nu-\epsilon^{\nu\bullet} P^\mu)
 -(\epsilon^{\mu\bullet}_\perp z^\nu -\epsilon^{\nu\bullet}_\perp z^\mu)\frac{M_T^2}{2 p z}\Bigg],\non \\
 {\cal T'^{\mu\nu}}&=&\frac{\epsilon^{\bullet\bullet} M_T^2}{2 (p z)^3} (p^\mu z^\nu-p^\nu z^\mu)
 ,\qquad\qquad
 {\cal U'^{\mu\nu}}=\frac{M_T^2}{p z} (\epsilon^\mu z^\nu-\epsilon^\nu
 z^\mu).
 \en
The derivations of Eqs. (\ref{Aphisigma}) and (\ref{Tphisigma}) are
shown in Appendix B. In contrast to Eqs. (\ref{Sphi}) $\sim$
(\ref{Tphi}), the twist-$2$ LCDAs do not disentangle entirely from
the higher twists in Eqs. (\ref{Aphisigma}) and (\ref{Tphisigma}).
Taking the product with $\epsilon_{\perp\mu} z_\nu$ and
$\epsilon_{\perp\mu\bullet} z_\nu$ in Eqs. (\ref{Aphisigma}) and
(\ref{Tphisigma}), respectively, to obtain,
 \be
 \langle 0 |[\bar q (-z)\sigma^{\mu\bullet}\epsilon_{\perp\mu}
  \gamma_5 q(z)]_2|A(P,\epsilon_{\lambda=\pm 1})\rangle &=& f^\perp_A \int^1_0 du
 \phi_{A\perp}(u) \frac{1}{2}(\epsilon\cdot \epsilon_\perp P z)\bigg[e^{i\xi p z}
 +\int^1_0 dte^{i\xi t^2 p z}\bigg],\non \\ \label{Aphisigmadis}\\
 \langle 0 |[\bar q (-z)\sigma^{\mu\bullet} \epsilon_{\perp\mu \bullet}
 q(z)]_2|T(P,\epsilon_{\lambda=\pm 1})\rangle
 &=& if^\perp_T M_T\int^1_0 du
 \phi_{T\perp}(u) \frac{1}{2} \epsilon^{\mu\bullet} \epsilon_{\perp \mu\bullet} \bigg[ e^{i\xi p z}
 +\int^1_0 dte^{i\xi t^2 p z}\bigg].\non \\ \label{Tphisigmadis}
 \en
Then, we use Eq. (\ref{twist2expandsigma}) and expand the right-hand
sides of Eqs. (\ref{Aphisigmadis}) and (\ref{Tphisigmadis}) as:
 \be
 &&~~~\sum^\infty_{n=0} \frac{1}{n!}\langle 0|\bar q
 (0)\frac{(n+1)(z\cdot \widehat{D})^n}{2 n+1}
 \sigma^{\mu\bullet}\epsilon_{\perp\mu} \gamma_5
 q(0)|A(P,\epsilon_{\lambda=\pm 1})\rangle \non \\
 &=& f^\perp_A  \sum^\infty_{n=0}
 \frac{i^n}{n!}\int^1_0 du \phi_{A\perp}(u) \frac{1}{2}(\epsilon\cdot \epsilon_\perp P z)
 (\xi p z)^n \Bigg[1+\int^1_0 dt t^{2n}\Bigg],\label{Ansigma}
 \en
 \be
 &&~~~\sum^\infty_{n=0} \frac{1}{n!}\langle 0|\bar q
 (0)\frac{(n+1)(z\cdot \widehat{D})^n}{2 n+1}
 \sigma^{\mu\bullet}\epsilon_{\perp\mu}  q(0)|T(P,\epsilon_{\lambda=\pm 1})\rangle\non \\
 &=& if^\perp_T M_T \sum^\infty_{n=0}
 \frac{i^n}{n!}\int^1_0 du \phi_{T\perp}(u) \frac{1}{2} \epsilon^{\mu\bullet} \epsilon_{\perp\mu \bullet}
 (\xi p z)^n \Bigg[1+\int^1_0 dt t^{2n}\Bigg], \label{Tnsigma}
 \en
Picking $n=0$ in Eqs. (\ref{Ansigma}) and (\ref{Tnsigma}), we
obtain:
 \be
 \langle 0|\bar q(0)\sigma^{\mu\bullet}\epsilon_{\perp\mu} \gamma_5
 q(0)|A(P,\epsilon_{\lambda=\pm 1})\rangle &=& f^\perp_A\int^1_0 du \phi_{A\perp}(u) (\epsilon\cdot
 \epsilon_\perp P z),\label{Ansigmafm}\\
 \langle 0|\bar q (0)\sigma^{\mu\bullet}\epsilon_{\perp\mu\bullet}  q(0)|T(P,\epsilon_{\lambda=\pm 1})\rangle
 &=& if^\perp_T M_T
 \int^1_0 du \phi_{T\perp}(u)  \epsilon^{\mu\bullet} \epsilon_{\perp\mu \bullet}. \label{Tnsigmafm}
 \en
It is worth noting that the author of Ref. \cite{Yang2} also
considered an approach that disentangled the twist-$2$ LCDAs from
the higher twists, in the case of an axial vector meson state
($\Gamma = \sigma^{\mu\nu}\gamma_5 $): Besides $z_\nu$, Eq.
(\ref{Aphisigma}) has taken the product with a term proportional to
$(\epsilon_\mu P z-P_\mu \epsilon z )$. We find this approach
equivalent to ours. The derivation is as follows. The term
$(\epsilon_\mu P z-P_\mu \epsilon z )$ can be expanded by using Eqs.
(\ref{Pp}) and (\ref{epsilon}) as
 \be
 \epsilon_\mu P z-P_\mu \epsilon z &=& p_\mu \frac{e z}{p z}P z
 -z_\mu \frac{e z M^2_A}{2 (p z)^2} P z + \epsilon_{\perp\mu} P z
 -p_\mu e z -z_\mu\frac{e z M^2_A}{2 p z}\non \\
 &=&  -z_\mu \frac{e z M^2_A}{p z}+\epsilon_{\perp\mu} P z.
 \en
The first term of last line has no contribution to the result
because $\sigma^{\mu\nu}$ is antisymmetric. 

\section{General Formulism in LFQM }
\subsection{Framework}

A meson bound state, consisting of a quark $q_1$ and an antiquark
$\bar q_2$ with total momentum $P$ and spin $J$, can be written as
(see, for example \cite{CCH1}):
 \be
 |M(P, ^{2S+1}L_J, J_z)\rangle =\int &&\{d^3k_1\}\{d^3k_2\} ~2(2\pi)^3
 \delta^3(\tilde P -\tilde k_1-\tilde k_2)~\non\\
 &&\times \sum_{\lambda_1,\lambda_2}
 \Psi^{JJ_z}_{LS}(\tilde k_1,\tilde k_2,\lambda_1,\lambda_2)~
 |q_1(k_1,\lambda_1) \bar q_2(k_2,\lambda_2)\rangle,\label{lfmbs}
 \en
where $k_1$ and $k_2$ are the on-mass-shell light-front momenta,
 \be
 \tilde k=(k^+, k_\bot)~, \quad k_\bot = (k^1, k^2)~,
 \quad k^- = \frac{m_q^2+k_\bot^2}{k^+},
 \en
and
 \be
 &&\{d^3k\} \equiv \frac{dk^+d^2k_\bot}{2(2\pi)^3}, \nonumber \\
 &&|q(k_1,\lambda_1)\bar q(k_2,\lambda_2)\rangle
 = b^\dagger_{\lambda_1}(k_1)d^\dagger_{\lambda_2}(k_2)|0\rangle,\\
 &&\{b_{\lambda'}(k'),b_{\lambda}^\dagger(k)\} =
 \{d_{\lambda'}(k'),d_{\lambda}^\dagger(k)\} =
 2(2\pi)^3~\delta^3(\tilde k'-\tilde k)~\delta_{\lambda'\lambda}.
 \nonumber
 \en
In terms of the light-front relative momentum variables $(u,
\kappa_\bot)$ defined by
 \be
 && k^+_1=(1-u) P^{+}, \quad k^+_2=u P^{+}, \nonumber \\
 && k_{1\bot}=(1-u) P_\bot+\kappa_\bot, \quad k_{2\bot}=u
 P_\bot-\kappa_\bot.
 \en
The relative momentum in $\hat{z}$ direction $\kappa_z$ can be
written as
 \be \label{eq:Mpz}
  \kappa_z=\frac{u M_0}{2}-\frac{m^2_2+\kappa^2_\perp}{2 u M_0}.
 \en

The momentum-space wave-function $\Psi^{JJ_z}_{LS}$ for a
$^{2S+1}L_J$ meson can be expressed as
 \be
 \Psi^{ JJ_z}_{LS}(\tilde k_1,\tilde k_2,\lambda_1,\lambda_2)
 = \frac{1}{\sqrt N_c}\la L S; L_z S_z|L S;J J_z\ra
 R^{SS_z}_{\lambda_1\lambda_2}(u,\kappa_\bot)~ \varphi_{LL_z}(u,
 \kappa_\bot),\label{Psi}
 \en
where $\varphi_{LL_z}(u,\kappa_\bot)$ describes the momentum
distribution of the constituent quarks in the bound state with the
orbital angular momentum $L$, $\langle L S; L_z S_z|L S;J
J_z\rangle$ is the corresponding Clebsch-Gordan coefficient and
$R^{SS_z}_{\lambda_1\lambda_2}$ constructs a state of definite spin
($S,S_z$) out of light-front helicity ($\lambda_1,\lambda_2$)
eigenstates. Explicitly,
 \be
 R^{SS_z}_{\lambda_1 \lambda_2}(u,\kappa_\bot)
 =\sum_{s_1,s_2} \langle \lambda_1|
  {\cal R}_M^\dagger(1-u,\kappa_\bot, m_1)|s_1\rangle
 \langle \lambda_2|{\cal R}_M^\dagger(u,-\kappa_\bot, m_2)
 |s_2\rangle \langle \frac{1}{2}\,\frac{1}{2};s_1
 s_2|\frac{1}{2}\frac{1}{2};SS_z\rangle,
 \en
where $|s_i\rangle$ are the usual Pauli spinors, and ${\cal R}_M$ is
the Melosh transformation operator~\cite{Jaus1}:
 \be
 \langle s|{\cal R}_M (u_i,\kappa_\bot,m_i)|\lambda\rangle
 &=&\frac{m_i+u_i M_0
 +i\vec \sigma_{s\lambda}\cdot\vec \kappa_\bot \times
 \vec n}{\sqrt{(m_i+u_i M_0)^2 + \kappa^{2}_\bot}},
 \en
with $u_1=1-u$, $u_2=u$, and 
$\vec n =(0,0,1)$ is a unit vector in the $\hat {z}$-direction. In
addition,
 \be
 M_0^2&=&(e_1+e_2)^2=\frac{m_1^2+\kappa^2_\bot}{u_1}+\frac{m_2^2+\kappa^2_\bot}{u_2},
 \\
 e_i&=&\sqrt{m^2_i+\kappa^2_\perp+\kappa^2_z}.\non
 \en
where $M_0$ is the invariant mass of $q\bar q$ and generally
different from the mass $M$ of meson which satisfies $M^2=P^2$. This
is due to the fact that the meson, quark and anti-quark cannot be
simultaneously on-shell. We normalize the meson state as
 \be
 \langle M(P',J',J'_z)|M(P,J,J_z)\rangle = 2(2\pi)^3 P^+
 \delta^3(\tilde P'- \tilde P)\delta_{J'J}\delta_{J'_z J_z}~,
 \label{wavenor}
 \en
in order that:
 \be
 \int \frac{du\,d^2\kappa_\bot}{2(2\pi)^3}~\varphi^{\prime*}_{L^\prime
 L^\prime_z}(u,\kappa_\bot)
 \varphi_{LL_z}(u,\kappa_\bot)
 =\delta_{L^\prime,L}~\delta_{L^\prime_z,L_z}.
 \label{momnor}
 \en
Explicitly, we have:
 \be
 \varphi_{1L_z}=\kappa_{L_z} \varphi_p,
 \en
where $\kappa_{L_z=\pm1}=\mp(\kappa_{\bot x}\pm i \kappa_{\bot
y})/\sqrt2$, $\kappa_{L_z=0}=\kappa_{z}$ are proportional to the
spherical harmonics $Y_{1L_z}$ in momentum space, $\varphi_p$ is the
distribution amplitude of $p$-wave meson. In general, for any
function $F(\vec{\kappa})$, $\varphi_p(u)$ has the form of:
 \be
 \varphi_p(u)=N \sqrt{\frac{d\kappa_z}{du}}F(\vec{\kappa}),\label{F}
 \en
where the normalization factor $N$ is determined from
Eq.~(\ref{momnor}).


In the case of a $p$-wave meson state, it is more convenient to use
the covariant form of $R^{SS_z}_{\lambda_1\lambda_2}$
\cite{Jaus1,CCH2,cheung}:
 \be
 \langle 1 S; L_z S_z|1 S;J J_z\rangle\, k_{L_z} \,R^{SS_z}_{\lambda_1\lambda_2}(u,\kappa_\bot)
 &=&\frac{\sqrt{k_1^+ k_2^+}}{\sqrt2~{\widetilde M_0}(M_0+m_1+m_2)}\non \\
 &&\times\bar u(k_1,\lambda_1)(\not\!\!\bar P+M_0)\Gamma_{^{2S+1}\!P_J}
 v(k_2,\lambda_2), \label{covariantp}
 \en
where
 \be
 &&\widetilde M_0\equiv\sqrt{M_0^2-(m_1-m_2)^2},\qquad\quad \bar
 P\equiv k_1+k_2,\non \\
 &&\bar u(k,\lambda) u(k,\lambda')=\frac{2
 m}{k^+}\delta_{\lambda,\lambda'},\qquad\quad \sum_\lambda u(k,\lambda)
 \bar u(k,\lambda)=\frac{\not\!k +m}{k^+},\non \\
 &&\bar v(k,\lambda) v(k,\lambda')=-\frac{2
 m}{k^+}\delta_{\lambda,\lambda'},\qquad\quad \sum_\lambda v(k,\lambda)
 \bar v(k,\lambda)=\frac{\not\!k -m}{k^+}.
 \en
For the scalar, axial-vector, and tensor mesons, we have:
 \be
 \Gamma_{^3\!P_0}&=&\frac{1}{\sqrt3}\left(\not\!\!K-\frac{K\cdot
 \bar P}{M_0}\right),\non\\
 \Gamma_{^1\!P_1}&=&-\epsilon\cdot K \gamma_5,\non\\
 \Gamma_{^3\!P_1}&=&\frac{1}{\sqrt2}\left((\not\!\!K-\frac{K\cdot
 \bar P}{M_0})\not\!\epsilon-\epsilon\cdot K\right)\gamma_5, \non\\
 \Gamma_{^3\!P_2}&=& \epsilon_{\mu\nu}\gamma^\mu(- K^\nu),\label{Gamma}
 \en
where $K\equiv (k_2-k_1)/2$ and:
 \be
 &&\epsilon^\mu_{\lambda=\pm 1} =
 \left[\frac{2}{ P^+} \vec \epsilon_\bot (\pm 1) \cdot
 \vec P_\bot,\,0,\,\vec \epsilon_\bot (\pm 1)\right],\non \\
 &&\vec \epsilon_\bot (\pm 1)=\mp(1,\pm i)/\sqrt{2}, \non\\
 &&\epsilon^\mu_{\lambda=0}=\frac{1}{M_0}\left(\frac{-M_0^2+P_\bot^2}{
 P^+},P^+,P_\bot\right).   \label{polcom}
 \en
Note that the polarization tensor of a tensor meson satisfies the
relations: $\epsilon_{\mu\nu}=\epsilon_{\nu\mu}$ and
$\epsilon_{\mu\nu}\bar P^\mu = \epsilon^\mu_\mu = 0$. Eqs.
(\ref{covariantp}) and (\ref{Gamma}) can be further reduced by the
applications of equations of motion on spinors:
 \be
 \langle 1 S; L_z S_z|1 S;J J_z\rangle\, k_{L_z}
 \,R^{SS_z}_{\lambda_1\lambda_2}(u,\kappa_\bot)
 =\frac{\sqrt{k_1^+ k_2^+}}{\sqrt2~{\widetilde M_0}}
 \bar u(k_1,\lambda_1)\Gamma'_{^{2S+1}\!P_J}
 v(k_2,\lambda_2), \label{covariantfurther}
 \en
where
 \be
 \Gamma'_{^3\!P_0}&=&-\frac{\widetilde M_0^2}{2\sqrt{3} M_0},\non\\
 \Gamma'_{^1\!P_1}&=&-\epsilon\cdot K \gamma_5,\non\\
 \Gamma'_{^3\!P_1}&=&\frac{-1}{2\sqrt{2} M_0}\left(\not\!\epsilon
 \widetilde M_0^2-2\epsilon\cdot K(m_1-m_2)\right)\gamma_5, \non\\
 \Gamma'_{^3\!P_2}&=& \epsilon_{\mu\nu}\left(\gamma^\mu+\frac{2 K^\mu}
 {M_0+m_1+m_2}\right)(- K^\nu).\label{Gammap}
 \en

\subsection{Analysis of Leading twist LCDAs}
Next, the matrix elements of Eqs. (\ref{S0}), (\ref{A0}),
(\ref{T1}), (\ref{Ansigmafm}), and (\ref{Tnsigmafm}) will be
calculated within LFQM, and the relevant leading twist LCDAs are
extracted. For the scalar meson state, we substitute Eqs.
(\ref{lfmbs}), (\ref{Psi}), and (\ref{covariantfurther}) into Eq.
(\ref{S0}) to obtain:
 \be
 \langle 0|\bar q_2 \gamma^\mu q_1|S(P)\rangle&=&N_c\int\{d^3
 k_1\}\sum_{\lambda_1,\lambda_2}\Psi_{LS}^{JJ_z}(k_1,k_2,\lambda_1,\lambda_2)
 \langle 0 |\bar q_2 \gamma^\mu q_1|q_1\bar q_2\rangle\non \\
 &=& -\sqrt{N_c}\int\{d^3 k_1\}\frac{\sqrt{k_1^+ k_2^+}}{\sqrt2~{\widetilde
 M_0}}\varphi_p{\rm Tr}\bigg[\gamma^\mu \Bigg(\frac{\not\!k_1+m_1}{k_1^+}\Bigg)\frac{\widetilde M_0^2}{2\sqrt{3}
 M_0}\Bigg(\frac{-\not\!k_2+m_2}{k^+_2}\Bigg)\bigg]\non \\
 &=&f_S P^\mu \int du \phi(u).
 \en
For the "good" component, $\mu=+$, the leading twist LCDA $\phi_S$
can be extracted as:
 \be
 \phi_S(u) = \frac{\sqrt{2 N_c}}{f_S}\int \frac{d^2
 \kappa_\perp}{2(2\pi)^3}\frac{[(1-u) m_2-u m_1]\widetilde
 M_0}{\sqrt{3u(1-u)}M_0}\varphi_p(u,\kappa_\perp).\label{Su}
 \en
In the quarkonium case $m_1=m_2=m$, Eq. (\ref{Su}) can be further
reduced as:
 \be
 \phi_S(u) = \frac{\sqrt{2}}{f_S}\int \frac{d^2
 \kappa_\perp}{2(2\pi)^3}\frac{(1-2 u) m}{\sqrt{u(1-u)}}\varphi_p(u,\kappa_\perp).\label{Sk}
 \en
A similar process can be used for the axial vector and tensor mesons
which correspond to Eqs. (\ref{A0}), (\ref{Ansigmafm}) and
(\ref{T1}), (\ref{Tnsigmafm}), respectively, and the leading twist
LCDAs are extracted as
 \be
 \phi_{^3\!A_1 \|}(u)&=&\frac{2\sqrt{3}}{f_{^3\!A_1}}\int \frac{d^2
 \kappa_\perp}{2(2\pi)^3}\frac{\kappa^2_\perp}{\sqrt{u(1-u)} M_0}\varphi_p(u,\kappa_\perp),\\
 \phi_{^1\!A_1\|}(u) &=& \frac{\sqrt{6}}{f_{^1\!A_1}}\int \frac{d^2
 \kappa_\perp}{2(2\pi)^3}\frac{(1-2 u)m}{\sqrt{u(1-u)}}\varphi_p(u,\kappa_\perp),\\
 \phi_{T\|}(u) &=& \frac{\sqrt{6}}{f_{T}}\int \frac{d^2
 \kappa_\perp}{2(2\pi)^3}\frac{(1-2 u)}{\sqrt{u(1-u)}}\left[M_0-m-\frac{\kappa^2_\perp}
 {M_0+2m}\right]\varphi_p(u,\kappa_\perp), \label{Tkperp}\\
 \phi_{^3\!A_1 \perp}(u)&=&\frac{\sqrt{3}}{f^\perp_{^3\!A_1}}\int \frac{d^2
 \kappa_\perp}{2(2\pi)^3}\frac{(1-2 u)m}{\sqrt{u(1-u)}}\varphi_p(u,\kappa_\perp), \\
 \phi_{^1\!A_1 \perp}(u)&=&\frac{\sqrt{6}}{f^\perp_{^1\!A_1}}\int \frac{d^2
 \kappa_\perp}{2(2\pi)^3}\frac{\kappa^2_\perp}{\sqrt{u(1-u)}M_0}\varphi_p(u,\kappa_\perp), \\
 \phi_{T\perp}(u) &=& \frac{\sqrt{6}}{f^\perp_{T}}\int \frac{d^2
 \kappa_\perp}{2(2\pi)^3}\frac{(1-2 u)}{\sqrt{u(1-u)}}\left[m+\frac{2\kappa^2_\perp}
 {M_0+2m}\right]\varphi_p(u,\kappa_\perp).\label{Tk}
 \en
From the normalization Eq. (\ref{normal}), some relations among the
constants $f_M$s could be easily obtained as:
 \be
 \sqrt{3}f_S=f_{^1\!A_1}=\sqrt{2}f^\perp_{^3\!A_1}\equiv f_{\rm odd},
 \quad {\rm and} \quad \frac{f_{^3\!A_1}}{\sqrt{2}}=f^\perp_{^1\!A_1}\equiv f_{\rm even},\label{indf}
 \en
then, the relations among the relevant LCDAs are:
 \be
 \phi_S=\phi_{^1\!A_1 \|}=\phi_{^3\!A_1\perp}\equiv \phi_{\rm odd},
 \quad {\rm and} \quad \phi_{^3\!A_1\|}=\phi_{^1\!A_1\perp}\equiv \phi_{\rm even},\label{indphi}
 \en
where the subscript "odd (even)" means the odd (even) function of
$u$ and
 \be
 \phi_{\rm odd}(u)&=& \frac{\sqrt{6}}{f_{\rm odd}}\int \frac{d^2
 \kappa_\perp}{2(2\pi)^3}\frac{(1-2 u)
 m}{\sqrt{u(1-u)}}\varphi_p(u,\kappa_\perp),\label{phioddm}\\
 \phi_{\rm even}(u)&=& \frac{\sqrt{6}}{f_{\rm even}}\int \frac{d^2
 \kappa_\perp}{2(2\pi)^3}\frac{\kappa^2_\perp}{\sqrt{u(1-u)} M_0}\varphi_p(u,\kappa_\perp)
 \label{phieven}
 \en
Note that Eqs. (\ref{indf}) and (\ref{indphi}) are independent of
the form of $\varphi_p$. Furthermore, in the nonrelativistic
situation, the momenta $\kappa_{z,\perp}$ are much smaller than
quark mass $m$, and $M_0$ can be reduced to approximately $2 m$.
Thus, we have:
 \be
 f_{\rm odd}\simeq f_T \simeq f^\perp_T,
 \quad\phi_{\rm odd}\simeq \phi_{T\|}\simeq \phi_{T\perp}.\label{almostsame}
 \en
and
 \be
 \phi_{\rm odd}(u)&\simeq& \frac{\sqrt{6}}{f_{\rm odd}}\int \frac{d^2
 \kappa_\perp}{2(2\pi)^3}\frac{(1-2 u)
 }{\sqrt{u(1-u)}}\frac{M_0}{2}\varphi_p(u,\kappa_\perp).\label{phiodd}
 \en
From Eqs. (\ref{phieven}) and (\ref{phiodd}), one can obtain:
 \be
 f_{\rm odd} \simeq f_{\rm even},\label{evenodd}
 \en
and relate the $\xi$-moments of the $\phi_{\rm even}$ to those of
the $\phi_{\rm odd}$ as
 \be
 \langle \xi^n\rangle_{\phi_{\rm even}}\simeq\frac{\langle \xi^{n+1}\rangle_{\phi_{\rm
 odd}}}{n+1},\label{ximoment}
 \en
with the function $F=F(|\vec{\kappa}|)$. The derivations of Eqs.
(\ref{evenodd}) and (\ref{ximoment}) are shown in Appendix C. The
above results are consistent with those of \cite{Bragutanew} in the
nonrelativistic approximation.\footnote{For the tensor mesons, Ref.
\cite{Bragutanew} has the relation $f_T = \sqrt{\frac{2}{3}}
f^\perp_T$. The additional $\sqrt{\frac{2}{3}}$ factor is the
Clebsch-Gordan coefficient of the polarization tensor
$\epsilon_{\mu\nu}$ for the tensor meson state
$T(P,\epsilon_{\lambda=0})$. This distinction is from the different
definition of $\phi_{T\|}$: There is a $\epsilon_{\mu\nu}$ in both
hand sides of Eq. (\ref{TL}). By contrast, in the upper part of Eq.
(4) of Ref. \cite{Bragutanew}, $\epsilon_{\mu\nu}$ appears only in
the left hand side.}

Next, we choose a Gaussian-like wave function, as shown in
\cite{CCH2}:
 \be
 \varphi_p(u,\kappa_\perp)=\frac{4\sqrt{2}}{\beta}\bigg(\frac{\pi}
 {\beta^2}\bigg)^{3/4}\sqrt{\frac{d\kappa_z}{du}}{\rm
 exp}\bigg(-\frac{|\vec \kappa|^2}{2 \beta^2}\bigg),\label{Gaussian1p}
 \en
for further calculations. In Eqs. (\ref{phioddm}) and
(\ref{phieven}), the $\kappa_\perp$ integrations can be performed as
follows:
 \be
 \phi_{\rm odd}(u) &=&\frac{\sqrt{6}(1-2 u)
 m}{f_{\rm odd}}\Bigg(\frac{2}{\pi}\Bigg)^{5/4}e^d\Gamma\left[\frac{5}{4},w\right],\label{oddgamma}\\
 \phi_{\rm even}(u) &=&\frac{\sqrt{3}u(1-u)\beta}{f_{\rm even}}\Bigg(\frac{2}{\pi}\Bigg)^{5/4}
 e^d\Bigg\{w\Gamma\left[-\frac{1}{4},w\right]+3
 \Gamma\left[\frac{3}{4},w\right]\Bigg\},\label{evengamma}
 \en
where $w= d/[4 u (1-u)]$, $d=m^2/2\beta^2$ and:
 \be
 \Gamma[a,w]=\int^\infty_w t^{a-1}e^{-t} dt\non
 \en
is the incomplete Gamma function. The incomplete gamma function may
be expressed quite elegantly in terms of the confluent
hypergeometric function:
 \be
 \Gamma[a,w]=\Gamma[a]-a^{-1}w^a\times~_1F_1(a;a+1;-w),\label{hpgeo}
 \en
where
 \be
 _iF_j(a_1,a_2,...,a_i;a'_1,a'_2,...,a'_j;w)=\sum^\infty_{n=0}\frac{(a_1)_n(a_2)_n
 ...(a_i)_n}{(a'_1)_n(a'_2)_n...(a'_j)_n}\frac{w^n}{n!},\label{1F1}
 \en
and $(a)_n=(a+n-1)!/(a-1)!$ is the Pochhammer symbol. However, the
$\kappa_\perp$ integrals in Eqs. (\ref{Tkperp}) and (\ref{Tk})
cannot be analytically performed. The crux is the term proportional
to $1/M_0+2 m$. This term may be rewritten and expanded as:
 \be \label{vwapp}
 \frac{\kappa^2_\perp}{M_0+2m}=\frac{\kappa^2_\perp}{4m\left(1+\frac{M_0-2m}{4m}\right)}
 =
 \frac{\kappa^2_\perp}{4m}\left[1-\left(\frac{M_0-2m}{4m}\right)+...\right].
 \en
We only consider the first two terms in the square bracket of Eq.
(\ref{vwapp}) because $M_0-2m$ goes to zero in the non-relativistic
approximation. Then, the approximate form of, for example,
$\phi_{T\|}(u)$ which containing the first one term and first two
terms in the square bracket of Eq. (\ref{vwapp}) are defined as:
 \be
 \phi_{T\|}^{(1)}(u) &=& \frac{\sqrt{6}}{f_{T}}\int \frac{d^2
 \kappa_\perp}{2(2\pi)^3}\frac{(1-2 u)}{\sqrt{u(1-u)}}\left[M_0-m-\frac{\kappa^2_\perp}
 {4m}\right]\varphi_p(u,\kappa_\perp), \label{Tkperp1}\\
 \phi_{T\|}^{(2)}(u) &=& \frac{\sqrt{6}}{f_{T}}\int \frac{d^2
 \kappa_\perp}{2(2\pi)^3}\frac{(1-2 u)}{\sqrt{u(1-u)}}\left[M_0-m-\frac{\kappa^2_\perp}
 {4m}\left(1-\frac{M_0-2 m}{4 m}\right)\right]\varphi_p(u,\kappa_\perp), \label{Tkperp2}
 \en
respectively. The $\kappa_\perp$ integrals can be performed as:
 \be
 \phi_{T\|}^{(1)}(u) &=& \frac{\sqrt{6}(1-2 u)m}{f_{T}}\Bigg(\frac{2}{\pi}\Bigg)^{5/4}e^d
 \Bigg\{-\frac{3}{4}\Gamma\left[\frac{5}{4},w\right]
 -\frac{1}{4 w}\Gamma\left[\frac{9}{4},w\right]+\frac{4 \beta}{\sqrt{2}m}\Gamma\left[\frac{7}{4},w\right]\Bigg\},
 \label{Tperpgamma}\\
 \phi_{T\|}^{(2)}(u) &=& \frac{\sqrt{6}(1-2 u)m}{f_{T}}\Bigg(\frac{2}{\pi}\Bigg)^{5/4}e^d
 \Bigg\{-\frac{5}{8}\Gamma\left[\frac{5}{4},w\right]
 -\frac{3}{8 w}\Gamma\left[\frac{9}{4},w\right]+\frac{15 \beta}
 {4\sqrt{2}m}\Gamma\left[\frac{7}{4},w\right]\non \\
 &&\qquad\qquad\qquad\qquad\qquad\quad+\frac{\beta}{4\sqrt{2}w m}
 \Gamma\left[\frac{11}{4},w\right]\Bigg\}.
 \label{Tperpgamma2}
 \en
We can find that the both curves of $\phi_{T\|}^{(1)}(u)$ and
$\phi_{T\|}^{(2)}(u)$ are almost overlap each other by using the
parameters in Table 1 (See Sec. IV). Therefore we only consider the
first term in the square bracket of Eq. (\ref{vwapp}) and take the
approximation $\phi_{T\|}(u)\simeq \phi_{T\|}^{(1)}(u)$. Finally the
form of $\phi_{T\perp}(u)$ is:
 \be
 \phi_{T\perp}(u) \simeq \frac{\sqrt{6}(1-2 u)m}{2 f^\perp_{T}}\Bigg(\frac{2}{\pi}\Bigg)^{5/4}e^d
 \Bigg\{\Gamma\left[\frac{5}{4},r\right]+
 \frac{1}{w}\Gamma\left[\frac{9}{4},w\right]\Bigg\}.\label{Tgamma}
 \en

In addition, the $\xi$-moments of Eqs. (\ref{oddgamma}),
(\ref{evengamma}), (\ref{Tperpgamma}), and (\ref{Tgamma}) can be
analytically expressed as:
 \be
 \langle \xi^{2l+1}\rangle_{\phi_{\rm odd}} &=& A_{\rm odd}
 \Bigg\{\frac{d^{5/4}
 \Gamma\left[\frac{-5}{4}\right]}{\Gamma\left[\frac{5}{4}+l\right]}~_1F_1\left(-\frac{1}{4}-l;
 \frac{9}{4};
 -d\right)+\frac{\Gamma\left[\frac{5}{4}\right]}{\Gamma\left[\frac{5}{2}+l\right]}~
 _1F_1\left(-\frac{3}{2}-l;-\frac{1}{4}; -d\right)\bigg\},\non \\
 \label{ximomentodd}\\
 \langle \xi^{2 l}\rangle_{\phi_{\rm even}} &=& A_{\rm even}\Bigg\{\frac{d^{3/4}\Gamma
 \left[\frac{1}{4}\right]}{\Gamma\left[\frac{7}{4}+l\right]}\Bigg[
 ~_1F_1\left(-\frac{3}{4}-l; \frac{3}{4};
 -d\right)-~_2F_2\left(-\frac{3}{4}-l,\frac{3}{4};-\frac{1}{4},\frac{7}{4};
 -d\right)\Bigg]\non \\
 &&\quad-\frac{3\Gamma\left[\frac{-1}{4}\right]}{4\Gamma\left[\frac{5}{2}+l\right]}
 ~_1F_1\left(-\frac{3}{2}-l;-\frac{3}{4}; -d\right)\Bigg\},\label{ximomenteven} \\
 \langle \xi^{2l+1}\rangle_{\phi_T}&\simeq& A_T\Bigg\{\frac{\sqrt{2}d^{5/4}\Gamma\left[\frac{-5}{4}\right]}
 {\Gamma\left[\frac{5}{4}+l\right]}\left[3~_1F_1\left(-\frac{1}{4}-l;\frac{9}{4};-d+\right)+\frac{5}{9}
 ~_2F_2\left(-\frac{1}{4}-l,\frac{9}{4};\frac{5}{4},\frac{13}{4};-d\right)\right]\non \\
 &+&\frac{\sqrt{2}\Gamma\left[\frac{5}{4}\right]}
 {\Gamma\left[\frac{7}{2}+l\right]}\left[\frac{5}{4 d}~
 _1F_1\left(-\frac{3}{2}-l;-\frac{5}{4};-d\right)
 +(9+4
 l)~_1F_1\left(-\frac{3}{2}-l;-\frac{1}{4};-d\right)\right]\non \\
 &-&16\frac{\beta}{m}\left[\frac{\Gamma\left[\frac{7}{4}\right]}{\Gamma\left[\frac{5}{2}+l\right]}
 ~_1F_1\left(-\frac{3}{2}-l;-\frac{3}{4};-d\right)+\frac{d^{7/4}\Gamma\left[-\frac{7}{4}\right]}
 {\Gamma\left[\frac{3}{4}+l\right]}~_1F_1\left(\frac{1}{4}-l;\frac{11}{4};-d\right)\right]\Bigg\},\non
 \\
 \label{ximomentT} \\
 \langle \xi^{2l+1}\rangle_{\phi_{T \perp}}&\simeq& A_{T\perp}\Bigg\{\frac{d^{5/4}
 \Gamma\left[-\frac{5}{4}\right]}{\Gamma\left[\frac{5}{4}+l\right]}\left[~_1F_1\left(-\frac{1}{4}-l;
 \frac{9}{4};-d\right)+\frac{5}{9}~_2F_2\left(-\frac{1}{4}-l,\frac{9}{4};\frac{5}{4},\frac{13}{4};-d
 \right)\right]\non \\
 &&\quad+\frac{2\Gamma\left[\frac{9}{4}\right]}{d\Gamma\left[\frac{5}{2}+l\right]}
 ~_1F_1\left(-\frac{3}{2}-l;-\frac{5}{4};-d\right)-\frac{4+2 l}{5+2 l}~_1F_1\left(-\frac{5}{2}-l;
 -\frac{5}{4};-d\right)\Bigg\},\label{ximomentTperp}
 \en
where
 \be
 A_{\rm odd} &=& \frac{\sqrt{6}m}{2f_{\rm odd}} \left(\frac{2}{\pi}\right)^{5/4}~e^d
 \Gamma\left[\frac{3}{2}+l\right],\quad\quad
 A_{\rm even} = \frac{\sqrt{3}\beta}{8f_{\rm even}} \left(\frac{2}{\pi}\right)^{5/4}~e^d
 \Gamma\left[\frac{1}{2}+l\right],\non \\
 A_{T} &=& -\frac{\sqrt{3}m}{8f_T} \left(\frac{2}{\pi}\right)^{5/4}~e^d
 \Gamma\left[\frac{3}{2}+l\right],\quad
 A_{T\perp} = \sqrt{\frac{3}{8}}\frac{m}{f_{T\perp}} \left(\frac{2}{\pi}\right)^{5/4}~e^d
 \Gamma\left[\frac{3}{2}+l\right],\non
 \en
and $l$ is a non-negative integer. The derivations of Eqs.
(\ref{ximomentodd}) $\sim$ (\ref{ximomentTperp}) used the formula:
 \be
 _1
 F_1(a;b;c)=\frac{b-1}{c}\left[~_1F_1(a;b-1;c)-~_1F_1(a-1;b-1;c)\right],
 \en
which is easily checked from the definition of the confluent
hypergeometric function Eq. (\ref{1F1}).

\section{Numerical results and discussions}
In this section, the LCDs, constants $f$s, and $\xi$-moments are
estimated. Prior to numerical calculations, the parameters $m$ and
$\beta$, which appeared in the wave function, must be determined
firstly. We consider the Hamiltonian of the $p$-wave heavy
quarkonium state as:
 \be
 H=2\sqrt{m^2+\vec\kappa^2}+b r-\frac{4\alpha_s}{3 r}+g_1 S \cdot
 L+g_2 S_{12}+g_3 s_1 \cdot s_2,\label{Ham}
 \en
where $b r$ $(-\frac{4\alpha_s}{3 r})$ is the linear (Coulomb)
potential, $S_{12}=(3 s_1 \cdot \hat r~s_2 \cdot \hat r - s_1\cdot
s_2)$ is the tensor force operator, and $g_{1,2,3}$ are the
functions of the relevant interquark potentials (the details are
shown in, for example, \cite{RR,BNS}). In this way, the mass
difference between the spin-single ground state $M(^1 P_1)$ and the
spin-weighted average of the triplet states $M(^3 P_J)\equiv[M(^3
P_0)+3 M(^3 P_1)+5 M(^3 P_2)]/9$ only has the contribution which
comes from the spin-spin interaction.\footnote{The calculations of
the expectation values of the fourth and fifth terms for the
Hamiltonian Eq. (\ref{Ham}) can refer to the appendix of Ref.
\cite{Jackson}.} Experimentally this hyperfine splitting is less
than $1$ Mev \cite{PDG08} in charmonium sector, and can be neglected
here. Then, we can use the mass $M(^3 P_J)$ and its variational
principle for the Hamiltonian Eq. (\ref{Ham}) in order to determine
parameters $m$ and $\beta$. In the process, the conjugate coordinate
wave function of Eq. (\ref{Gaussian1p})
 \be
 \tilde{\varphi}^{1P}_m(r)&=&\sqrt{\frac{8}{3}}\frac{\beta^{3/2}}{\pi^{1/4}}~\beta r~{\rm exp}
 \left(-\frac{\beta^2 r^2}{2}\right)Y_{1m}(\theta,\phi),\label{coordinate1P}
 \en
is necessary. The values of the additional parameters $b$ and
$\alpha_s$ come from literature \cite{book}:
 \be
 b=0.18~{\rm GeV}^2,\qquad \alpha_s = 0.36,\label{parameterab}
 \en
for the heavy quarkonium states. We individually vary $b$ and
$\alpha_s$ to realize how they connect to $m$ and $\beta$. The
results are shown in Table I. We find the parameters $m$ and $\beta$
insensitively depend on $b$ and $\alpha_s$.
\begin{table}[ht!]
\caption{\label{tab:parameter} The connections among the parameters
$b$, $\alpha_s$ and $m$, $\beta$ of the $p$-wave heavy quarkonium
states. }
\begin{ruledtabular}
\begin{tabular}{cc|cc|cc}
 $b({\rm GeV}^2)$ & $\alpha_s$ & $m_c({\rm GeV})$ & $\beta_{cc}({\rm GeV})$ & $m_b({\rm GeV})$ &  $\beta_{bb}({\rm
 GeV})$\\\hline
$0.18\pm 0.02$ & $0.36$   & $1.38^{-0.03}_{+0.04}$ &
$0.489^{+0.015}_{-0.014}$ & $4.76\mp 0.02$&  $0.791^{+0.022}_{-0.023}$\\
 $0.18$ & $0.36\pm 0.04$ & $1.38 \pm 0.01$ & $0.489\pm 0.007$& $4.76\pm 0.02$
 & $0.791^{+0.022}_{-0.021}$
\end{tabular}
\end{ruledtabular}
\end{table}

Next, we use the center values of the parameters in Table I to
calculate and plot the LCDAs in Eqs. (\ref{oddgamma}),
(\ref{evengamma}), (\ref{Tperpgamma}), and (\ref{Tgamma}). The
results are as follows:
 \be
 f_{\rm odd}=0.0884~ {\rm GeV},\qquad f_{\rm even}=0.109~ {\rm GeV},\non \\
 f_{T\|}=0.124~ {\rm GeV},\qquad f_{T\perp}=0.0978~ {\rm GeV},
 \en
for the charmonium states, and
 \be
 f_{\rm odd}=0.0674~ {\rm GeV},\qquad f_{\rm even}=0.0716~ {\rm GeV},\non \\
 f_{T\|}=0.0750~ {\rm GeV},\qquad f_{T\perp}=0.0692~ {\rm GeV},
 \en
for the bottomonium state. The curves of LCDAs are shown in Figs. 1
and 2.
 \begin{figure}
 \includegraphics*[width=4in]{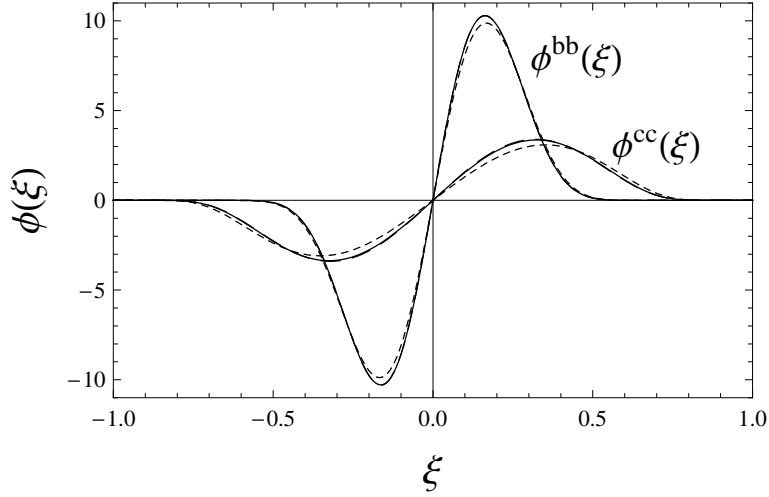}
 \caption{ The LCDAs $\phi_{\rm odd}(\xi)$ (solid
 line), $\phi_{T\|}(\xi)$ (dashed line), and $\phi_{T\perp}(\xi)$
 (long dashed line) of the charmonium and bottomonium states.
 The solid line completely overlaps the long dashed line.}
  \label{fig:phicb}
 \end{figure}
 \begin{figure}
 \includegraphics*[width=4in]{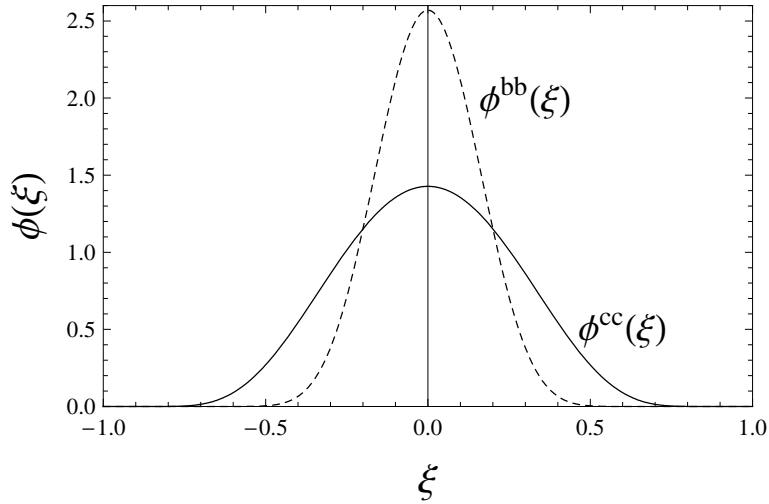}
 \caption{ The LCDAs $\phi_{\rm even}(\xi)$ of the charmonium (solid
 line) and bottomonium (dashed line) states.}
 \label{fig:evenbb}
 \end{figure}
These results are consistent with Eqs. (\ref{almostsame}) and
(\ref{evenodd}). 
In addition, the curve of $\phi^{bb}(\xi)$ in Fig. 2, in which $\xi$
is peaked around zero, was sharper than that of $\phi^{cc}(\xi)$.
This meant that the momentum fraction $u$ in the bottomonium state
is more centered on $1/2$ than in the charmonium state, which is
reasonable, as the mass of $b$ quark is larger than that of $c$
quark. A similar situation exists for the odd functions in Fig. 1.

Finally we show the LCDAs in terms of the $\xi$-moments. Eqs.
(\ref{ximomentodd}), (\ref{ximomentT}), (\ref{ximomentTperp}), and
(\ref{ximomenteven}) are calculated for $l=3,5,7$ and $l=2,4,6$,
respectively. The results, which compare with the other theoretical
evaluations, are as shown in Tables 2 and 3. In Table 2, Refs.
\cite{Bragutanew} used the QCD sum rules with the non-relativistic
wave functions. The authors of Ref. \cite{BT} calculated in the
framework of the Buchmuller-Tye potential model, and found their
results are in agreement with experiments which included the
leptonic widths and hyperfine splittings. The authors of
\cite{EGKLY} calculated in the Cornell potential. For the charmonium
sector, our results are not only consistent with those of
\cite{Bragutanew,BT,EGKLY}, but also conform to Eq.
(\ref{ximoment}).
\begin{table}[ht!]
\caption{\label{tab:cximoment} The $\xi$-moments for the $p$-wave
charmonium states. ($^\dag$ $\langle\xi^i\rangle_{\phi_{\rm
odd}}=\langle \xi^i\rangle_{h_c}$. $^\ddag$
$\langle\xi^i\rangle_{\phi_{\rm even}}=\langle
\xi^{i+1}\rangle_{h_c}/(i+1)$) }
\begin{ruledtabular}
\begin{tabular}{c|l|l|l|l}
 moment & this work & \cite{Bragutanew}$^{\dag\ddag}$ & \cite{BT} & \cite{EGKLY} \\ \hline
 $\langle \xi^3\rangle_{\phi_{\rm odd}}$ & $0.190$  & $0.18\pm 0.03$
 & $0.18$ &  $0.16$ \\
 $\langle \xi^5\rangle_{\phi_{\rm odd}}$ & $0.0507$ & $0.050\pm 0.010$
 & $0.047$ & $0.040$ \\
 $\langle \xi^7\rangle_{\phi_{\rm odd}}$ & $0.0164$& $0.017\pm 0.004$ & $0.016$ & $0.013$\\ \hline
  $\langle \xi^3\rangle_{\phi_T}$ & $0.206$  &  &  & \\
 $\langle \xi^5\rangle_{\phi_T}$ & $0.0583$ &  &  &\\
 $\langle \xi^7\rangle_{\phi_T}$ & $0.0198$ &  &  &\\ \hline
 $\langle \xi^3\rangle_{\phi_{T\perp}}$ & $0.188$  &  &  & \\
 $\langle \xi^5\rangle_{\phi_{T\perp}}$ & $0.0498$ &  &  & \\
 $\langle \xi^7\rangle_{\phi_{T\perp}}$ & $0.0160$ & &  & \\ \hline
 $\langle \xi^2\rangle_{\phi_{\rm even}}$ & $0.0662$  & $0.06\pm 0.01$ &  & \\
 $\langle \xi^4\rangle_{\phi_{\rm even}}$ & $0.0110$ &  $0.010\pm 0.002$&  & \\
 $\langle \xi^6\rangle_{\phi_{\rm even}}$ & $0.00261$ & $0.0024\pm 0.0006$& & \\
\end{tabular}
\end{ruledtabular}
\end{table}
\begin{table}[ht!]
\caption{\label{tab:bximoment} The $\xi$-moments for the $p$-wave
bottomonium states in this work. }
\begin{ruledtabular}
 \begin{tabular}{c|l||c|l||c|l||c|l}
 $\langle \xi^3\rangle_{\phi_{\rm odd}}$ & $0.0666$  & $\langle \xi^3\rangle_{\phi_T}$ & $0.0691$ &
 $\langle \xi^3\rangle_{\phi_{T_\perp}}$ & $0.0665$  & $\langle \xi^2\rangle_{\phi_{\rm even}}$ & $0.0226$\\
 $\langle \xi^5\rangle_{\phi_{\rm odd}}$ & $0.00685$ & $\langle \xi^5\rangle_{\phi_T}$ & $0.00735$ &
 $\langle \xi^5\rangle_{\phi_{T_\perp}}$ & $0.00684$ & $\langle \xi^4\rangle_{\phi_{\rm even}}$ & $0.00142$   \\
 $\langle \xi^7\rangle_{\phi_{\rm odd}}$ & $0.000922$&  $\langle \xi^7\rangle_{\phi_T}$ & $0.00102$ &
 $\langle \xi^7\rangle_{\phi_{T_\perp}}$ & $0.000919$ & $\langle \xi^6\rangle_{\phi_{\rm even}}$ & $0.000139$\\
\end{tabular}
\end{ruledtabular}
\end{table}
In the framework of NRQCD, the authors of Ref. \cite{Bragutanew}
related the relative velocity of quark-antiquark pair inside the
$p$-wave charmonium state to the $\xi$-moments as:
 \be
 \langle v^n \rangle_p=\frac{n+3}{3} \langle \xi^{n+1} \rangle + O(v^{n+2})
 \en
If the $\xi$-moments $\langle \xi^n \rangle_{\phi_{\rm odd}}$ are
considered, we obtain:
 \be
 \langle v^2 \rangle_p&=& 0.317,\non\\
 \langle v^4 \rangle_p&=& 0.118,\non \\
 \langle v^6 \rangle_p&=& 0.0492,\non
 \en
for the charmonium sector and
 \be
 \langle v^2 \rangle_p&=& 0.111,\non\\
 \langle v^4 \rangle_p&=& 0.0160,\non \\
 \langle v^6 \rangle_p&=& 0.00277,\non
 \en
for the bottomonium sector. These results are consistent with the
values $\langle v^2 \rangle_{cc}\approx 0.3$ and $\langle v^2
\rangle_{bb}\approx0.1$ used in NRQCD.
\section{Conclusions}
This study discussed the leading twist LCDAs of the $p$-wave heavy
quarkonium states within the light-front approach. The twist-$2$
LCDAs have been disentangled from the higher twists by appropriately
coping with the nonlocal operators $\bar q(-z)\Gamma q(z)$. For the
$\Gamma = \sigma_{\mu\nu}(\gamma_5)$ case, we proved that our method
is equivalent to that of Ref. \cite{Yang2}. Next, these LCDAs have
been shown in terms of the light-front variables $(u,\kappa_\perp)$
and the relevant decay constants. We found that the decay constants
and LCDAs had the following relations:
$\sqrt{3}f_S=f_{^1\!A_1}=\sqrt{2}f^\perp_{^3\!A_1}(\equiv f_{\rm
odd})$, $f_{^3\!A_1}/{\sqrt{2}}=f^\perp_{^1\!A_1}(\equiv f_{\rm
even})$, and $\phi_S=\phi_{^1\!A_1 \|}=\phi_{^3\!A_1\perp}(\equiv
\phi_{\rm odd})$, $\phi_{^3\!A_1\|}=\phi_{^1\!A_1\perp}(\equiv
\phi_{\rm even})$. If one takes the non-relativistic limit and the
wave function as a function of $|\vec \kappa|$, then the above
relations among the decay constants could be further simplified as
$f_{\rm odd}\simeq f_T\simeq f^\perp_T\simeq f_{\rm even}$, and in
addition, the $\xi$-moments of $\phi_{\rm odd}$ and $\phi_{\rm
even}$ have the relation: $\langle \xi^n\rangle_{\phi_{\rm
even}}=\langle \xi^n\rangle_{\phi_{\rm odd}}/(n+1)$.

The $\kappa_\perp$ integrations for the equations of LCDAs and
$\xi$-moments could be analytically performed when the Gaussian-type
wave function is considered. The parameters $m$ and $\beta$, which
appear in the wave function, were determined by taking the mass of
the spin-weighted average of the triplet state $M(^3 P_J)$ and the
variational principle for its Hamiltonian into account. We found the
parameters $m$ and $\beta$ insensitively depended on the linear
potential constant $b$ and the strong coupling constant $\alpha_s$.
The curves and the corresponding decay constants of the LCDAs
$\phi_{\rm odd}$, $\phi_{T\|}$, $\phi_{T\perp}$, and $\phi_{\rm
even}$ were plotted and calculated for the charmonium and
bottomonium states. These results are consistent with the relations
which are mentioned in last paragraph. However, the value of $f_{\rm
odd}$ is about a factor of two smaller than that in
\cite{Bragutanew} which was studied within QCD sum rules. In
addition, the first three $\xi$-moments were calculated, and were
consistent with those of other theoretical approaches. The relative
velocity of quark-antiquark pair $\langle v^2 \rangle$ of charmonium
and bottomonium states were also estimated and were consistent with
those used in NRQCD.


{\bf Acknowledgements}\\
The author would like to thank Kwei-Chou Yang and A. V. Luchinsky for their helpful
discussions. This work is supported in part by the National Science
Council of R.O.C. under Grant No NSC-96-2112-M-017-002-MY3.

\appendix
\section{Derivations of Eqs. (\ref{Sphi}) $\sim$ (\ref{Tphi}) }
Firstly, Eqs. (\ref{S}), (\ref{AL}), and (\ref{TL}) are rewritten as
 \be
 \langle 0|\bar q (z) \gamma_\mu q (-z)|S(P)\rangle &=& f_S \int^1_0 du~e^{i\xi p z}\left\{P_\mu
  \phi_S(u)+z_\mu \frac{M_S^2}{2p z}\left[g_S(u)-\phi_S(u)\right]\right\}, \label{a1}\\
 \langle 0|\bar q (z) \gamma_\mu \gamma_5 q (-z)|A(P,\epsilon_{\lambda=0})\rangle
 &=&if_A M_A \int^1_0 du~e^{i\xi p z}\Big\{\epsilon_\mu
 \phi_{A\|}(u) +\epsilon _{\perp \mu}\left[g_{A\perp}(u)-\phi_{A\|}(u)\right]\non \\
 &&~~~~~~~~~~~~~~~~~~~~~~- z_\mu \frac{\epsilon z}{2 (p z)^2}M^2_A
 \left[g_{A3}(u)-\phi_{A\|}(u)\right]\Big\}, \label{a2}\\
 \langle 0|\bar q (z) \gamma_\mu q (-z)|T(P,\epsilon_{\lambda=0})\rangle
 &=&f_T M_T^2 \int^1_0 du~e^{i\xi p z}\Big\{ \frac{\epsilon_{\mu\bullet}}{p z}
 \phi_{T\|}(u) +\frac{\epsilon _{\perp \mu\bullet}}{p z}\left[g_{T\perp}(u)-\phi_{T\|}(u)\right]\non \\
 &&~~~~~~~~~~~~~~~~~~~~~~- z_\mu \frac{\epsilon_{\bullet\bullet}}{2 (p z)^3}M^2_T
 \left[g_{T3}(u)-\phi_{T\|}(u)\right]\Big\},\label{a3}
 \en
respectively. Next, we sandwich both sides of Eq. (\ref{t1}) between
the vacuum and, for example, the scalar meson state
 \be
 \langle 0|[\bar q (-z) \gamma_\mu q(z)]_2 |S(P)\rangle &=& \int^1_0 dt
 \frac{\partial}{\partial z^\mu}\langle 0|\bar q (-t z) \not \!z q(t z)
 |S(P)\rangle \non \\
 &=&f_S\int^1_0 dt \frac{\partial}{\partial z^\mu}P z \int^1_0 du
 e^{i\xi p z} \phi_S(u) \non \\
 &=&f_S \int^1_0 du \phi_S(u)\Bigg\{P_\mu \int^1_0 dt e^{i\xi t p
 z}+ p_\mu (i \xi P z) \int^1_0 dt t e^{i\xi t p
 z}\Bigg\}.\label{a4}
 \en
The second term of the last line can be further calculated as:
 \be
 p_\mu (i \xi P z) \int^1_0 dt t e^{i\xi t p z} = p_\mu\frac{P z}{p
 z} \int^1_0 dt t \frac{\partial}{\partial t}e^{i\xi t p z}
 = p_\mu \left[e^{i\xi p z}-\int^1_0 d t e^{i\xi t p
 z}\right]. \label{a5}
 \en
We can substitute Eq. (\ref{a4}) for Eq. (\ref{a5}) and obtain Eq.
(\ref{Sphi}). In addition, the same process can be used to obtain
Eqs. (\ref{Aphi}) and (\ref{Tphi}). In fact, the above process has
been used for the vector meson state in Ref. \cite{Ball2}.

\section{Derivations of Eqs. (\ref{Aphisigma}) and (\ref{Tphisigma})}
Eqs. (\ref{AT}) and (\ref{TT}) can be rewritten as:
 \be
 \langle 0|\bar q (z) \sigma_{\mu\nu} \gamma_5 q (-z)|A(P,\epsilon_{\lambda=\pm 1})\rangle
 &=&f^\perp_A \int^1_0 du~e^{i\xi p z}\Big\{(\epsilon_\mu P_\nu-\epsilon_\nu P_\mu)
 \phi_{A\perp} (u)\non \\
 &+&(p_\mu z_\nu-p_\nu z_\mu)\frac{M^2_A \epsilon z}{(p
 z)^2}[h_{A\|}(u)-\phi_{A\perp} (u)]\non \\
 &+&(\epsilon_{\perp\mu} z_\nu-\epsilon_{\perp\nu} z_\mu) \frac{M^2_A}{2 p z}
 [h_{A3}(u)-\phi_{A\perp} (u)]\Big\},\label{b1}\\
 \langle 0|\bar q (z) \sigma_{\mu\nu} q (-z)|T(P,\epsilon_{\lambda=\pm 1})\rangle
 &=&if^\perp_T M_T \int^1_0 du~e^{i\xi p z}\Big\{\frac{(\epsilon_{\mu\bullet} P_\nu-\epsilon_{\bullet\nu}
 P_\mu)}{p z}
 \phi_{T\perp} (u)\non \\
 &+&(p_\mu z_\nu-p_\nu z_\mu)\frac{M^2_T \epsilon_{\bullet\bullet}}{(p
 z)^3}[h_{T\|}(u)-\phi_{T\perp} (u)]\non \\
 &+&(\epsilon_{\perp\mu\bullet} z_\nu-\epsilon_{\perp\bullet\nu} z_\mu) \frac{M^2_T}{2 (p z)^2}
 [h_{T3}(u)-\phi_{T\perp} (u)]\Big\},\label{b2}
 \en
respectively. Then, we sandwich both sides of Eq. (\ref{t2}) between
the vacuum and, for example, the axial-vector meson state,
 \be
 &&\langle 0|[\bar q (-z) \gamma_\mu \gamma_5 q(z)]_2
 |A(P,\epsilon_{\lambda=\pm 1})\rangle\non \\
 &=& \int^1_0 dt \Bigg[\frac{\partial}
 {\partial z^\mu} \langle 0|\bar q (-t^2 z) \sigma_{\bullet\nu} \gamma_5 q(t^2
 z)|A(P,\epsilon)\rangle +z^\alpha \frac{\partial}
 {\partial z^\nu} \langle 0|\bar q (-t^2 z) \sigma_{\mu\alpha} \gamma_5 q(t^2 z)
 |A(P,\epsilon)\rangle\Bigg] \non \\
 &=& f_A^\perp \int^1_0 du \Bigg\{\phi_{A\perp}(u)\Bigg[(\epsilon_\mu P_\nu
 -\epsilon_\nu P_\mu)\int^1_0 dt e^{i\xi t^2 p z}+2 p z {\cal S_{\mu\nu}}
 (i\xi)\int^1_0 dt t^2 e^{i\xi t^2 p z}\Bigg]\non \\
 &&\qquad\qquad+\big(h_{A\|}(u)-\phi_{A\perp}(u)\big)\Bigg[{\cal U_{\mu\nu}}\int^1_0dte^{i\xi t^2 p z}+
 2 p z {\cal T_{\mu\nu}}(i\xi)\int^1_0 dt t^2e^{i\xi t^2 p
 z}\Bigg]\bigg\}. \label{b3}
 \en
We can further calculate the integral as:
 \be
 i \xi \int^1_0 dt t^2 e^{i\xi t^2 p z} = \frac{1}{2 p
 z} \int^1_0 dt t \frac{\partial}{\partial t}e^{i\xi t^2 p z}
 =\frac{1}{2 p z} \left[e^{i\xi p z}-\int^1_0 d t e^{i\xi t^2 p
 z}\right], \label{b4}
 \en
and then substitute Eq. (\ref{b3}) for Eq. (\ref{b4}) to obtain Eq.
(\ref{Aphisigma}). The same process can be used to obtain Eq.
(\ref{Tphisigma}).

\section{Derivations of Eqs. (\ref{evenodd}) and (\ref{ximoment})}
From Eqs. (\ref{phieven}),
(\ref{phiodd}) and the normalization Eq. (\ref{normal}), we have
 \be
 f_{\rm odd} &\simeq& \sqrt{6}\int^1_{-1}d\xi \xi\int\frac{d^2\kappa_\perp}{2 (2\pi)^3}
 \frac{\xi}{\sqrt{1-\xi^2}}\frac{M_0}{2}\varphi_p(\xi,\kappa_\perp),\label{odd1}\\
 f_{\rm even} &=& \sqrt{6}\int^1_{-1}d\xi \int\frac{d^2\kappa_\perp}{2 (2\pi)^3}
 \frac{1}{\sqrt{1-\xi^2}}\frac{\kappa^2_\perp}{M_0}\varphi_p(\xi,\kappa_\perp)\label{even1}.
 \en
Taking Eqs. (\ref{odd1}) and (\ref{even1}) integration by parts with
respect to $\kappa^2_\perp$ and $\xi$, respectively, and using
normalization conditions: $\varphi_p(\xi,\kappa_\perp)$ must go to
zero when $\xi$ ($\kappa_\perp$) go to $\pm 1$ (infinity), and thus,
we can obtain:
 \be
 f_{\rm odd} &\simeq& -\sqrt{6}\int^1_{-1}d\xi \xi\int\frac{d^2\kappa_\perp}{2 (2\pi)^3}
 \kappa^2_\perp \frac{\xi}{\sqrt{1-\xi^2}}\frac{d}{d\kappa^2_\perp}\left[\frac{M_0}{2}
 \varphi_p(\xi,\kappa_\perp)\right],\label{odd2}\\
 f_{\rm even} &=& -\sqrt{6}\int^1_{-1}d\xi \xi\int\frac{d^2\kappa_\perp}{2 (2\pi)^3}
 \kappa^2_\perp \frac{d}{d\xi}\left[\frac{1}{\sqrt{1-\xi^2}}\frac{1}{M_0}
 \varphi_p(\xi,\kappa_\perp)\right]\label{even2}.
 \en
From Eq. (\ref{F}), the differentiations in Eqs. (\ref{odd2}) and
(\ref{even2}) can be expanded as:
 \be
 \frac{\xi}{\sqrt{1-\xi^2}}\frac{d}{d\kappa^2_\perp}\left[\frac{M_0}{2}
 \varphi_p(\xi,\kappa_\perp)\right]&=&\frac{3\xi}{2\sqrt{M_0}(1-\xi^2)^2}F(\vec
 \kappa)+\frac{\xi M_0^{3/2}}{2 (1-\xi^2)}\frac{d}{d\kappa^2_\perp}F(\vec
 \kappa),\label{odd3}\\
 \frac{d}{d\xi}\left[\frac{1}{\sqrt{1-\xi^2}}\frac{1}{M_0}
 \varphi_p(\xi,\kappa_\perp)\right]&=&\frac{3\xi}{2\sqrt{M_0}(1-\xi^2)^2}F(\vec
 \kappa)+\frac{1}{\sqrt{M_0}(1-\xi^2)}\frac{d}{d\xi}F(\vec
 \kappa),\label{even3}
 \en
respectively. If the function $F=F(|\vec \kappa|)$, one can expand
the arbitrary function $F(|\vec \kappa|)$ as a polynomial of $|\vec
\kappa|$:
 \be
 F(|\vec \kappa|) = \sum_{s}c_s |\vec \kappa|^s.
 \en
Using the relation $|\vec \kappa|=\sqrt{M_0^2/4-m^2}$, one can
calculate the differentiations in Eqs. (\ref{odd3}) and
(\ref{even3}) as:
 \be
 \frac{\xi M_0^{3/2}}{2 (1-\xi^2)}\frac{d}{d\kappa^2_\perp}F(|\vec
 \kappa|)=\frac{1}{\sqrt{M_0}(1-\xi^2)}\frac{d}{d\xi}F(|\vec
 \kappa|)=\sum_s c_s \frac{\xi s
 M_0^{3/2}}{4(1-\xi^2)^2}\left(\frac{M_0^2}{4}-m^2\right)^{s/2-1}.
 \en
Then, Eq. (\ref{evenodd}) is obtained. Regarding the $\xi$-moments
of $\phi_{\rm odd}$ and $\phi_{\rm even}$, we have:
 \be
 \langle \xi^{n+1} \rangle_{\phi_{\rm odd}}&\simeq& \frac{\sqrt{6}}{f_{\rm odd}}\int^1_{-1}d\xi
 \xi^{n+1}\int\frac{d^2\kappa_\perp}{2 (2\pi)^3}
 \frac{\xi}{\sqrt{1-\xi^2}}\frac{M_0}{2}\varphi_p(\xi,\kappa_\perp),\label{odd4}\\
 \langle \xi^{n} \rangle_{\phi_{\rm even}} &=& \frac{\sqrt{6}}{f_{\rm even}}
 \int^1_{-1}d\xi \xi^n \int\frac{d^2\kappa_\perp}{2 (2\pi)^3}
 \frac{1}{\sqrt{1-\xi^2}}\frac{\kappa^2_\perp}{M_0}\varphi_p(\xi,\kappa_\perp)\label{even4}.
 \en
Using the above processes, Eq. (\ref{ximoment}) can be obtained.

\end{document}